# Non-Hermitian Anderson Transport

Sebastian Weidemann[1,*], Mark Kremer[1,*], Stefano Longhi[2,3] and Alexander Szameit[1]

Institute of Physics, University Rostock, Albert-Einstein-Straße 23, 18059 Rostock, Germany.
[2]Dipartimento di Fisica, Politecnico di Milano, Piazza Leonardo da Vinci 32, 20133 Milano, Italy.
[3]FISC (UIB-CSIC), Instituto de Fisica Interdisciplinar y Sistemas Complejos - Palma de Mallorca, Spain.
*These authors contributed equally to this work.

**Anderson's groundbreaking discovery that the presence of stochastic imperfections in a crystal may result in a sudden breakdown of conductivity[1] revolutionized our understanding of disordered media. After stimulating decades of lively studies[2], Anderson localization has found intriguing applications in various areas of physics, such mesoscopic physics[3], strongly-correlated systems[4], light localization[5,6], cavity quantum electrodynamics[7], random lasers[8,9] and topological phases of matter[10–12]. However, a fundamental assumption in Anderson's treatment is that no energy is exchanged with the environment, in contrast to the common knowledge that every real system is subject to dissipation. Recently, a growing number of theoretical studies has addressed disordered media with dissipation[13–21]. In particular it has been predicted that in such systems all eigenstates exponentially localize[13], similar to the original case without dissipation that Anderson considered. However, in dissipative systems an eigenstate analysis is insufficient for characterizing the transport dynamics of wave, in stark contrast to Hermitian systems, where the localization of all eigenstates necessarily suppresses transport. In our work, we show in theory and experiment that systems with dissipative disorder allow for a new type of spatial transport, despite the fact that all eigenstates are exponentially localized. This "Anderson transport" is characterized by super-diffusive spreading and ultra-far spatial jumps between localized states. We anticipate our findings to mark the starting point towards novel phenomena in dissipative media, which are subject to general non-Hermitian disorder.**

In the famous Drude model[22], electronic conductivity relies on the idea that moving electrons, understood as particles, are decelerated by random collisions with the positive ions of the crystalline lattice, resulting in diffusive transport (Fig. 1a). However, the discovery of quantum mechanics showed that in solids electrons exhibit wave characteristics[23]. Consequently, Anderson included interference in the treatment of electron dynamics and discovered an unexpected metal–insulator transition, that is, lattice disorder can feature a sudden breakdown of electron transport[1] (Fig. 1b). Being general in its setting, Anderson's model also applies for systems beyond electrons in solids[24,25], such as photonics[5,6], matter waves[26,27], and sounds waves[28]. Both theories of Drude and Anderson naturally assume conservation of energy within the system. Conversely, the majority of systems are dissipative and exchange energy with their environment. This poses a fundamental question: How wave transport is affected by the presence of disorder *and* dissipation?

In order to address this open problem, dissipative (i.e. non-Hermitian) manifestations of disordered systems have been proposed and theoretically analyzed[14–21,29]. These works suggest that the presence of tailored dissipation can have various outcomes depending on the tailoring itself, such as suppressed diffusion[20] or tuneable localization of states[16]. However, in particular the case of fully stochastic dissipation is of fundamental interest, where both real and imaginary parts of the potential exhibit random disorder. This represents a stochastic energy exchange of the system with the environment and, as such, constitutes the direct non-Hermitian analogue[13] to the conventional Anderson model.

It has been numerically predicted[13,19] that disorder in the imaginary part of a potential leads to exponential localization of all eigenstates, like in the Hermitian Anderson case. This suggests that disorder-induced localization of eigenstates does not depend on whether the disorder is imposed onto the real or the imaginary part of the potential. In a Hermitian system, exponential localization of all eigenstates necessarily hinders transport, such that any spatial dynamics is suppressed. This is a well-known consequence of Anderson localization. In dissipative systems, however, this chain of cause-and-effect does not hold anymore, such that the physics of wave dynamics in a system with stochastic dissipation remains an exciting, yet open question.

We show, in theory and experiment, that systems with stochastic dissipation exhibit transport, even though all eigenstates are exponentially localized – which is in stark contrast to the common intuition for conservative systems. We address this issue by formulating a theoretical model, which predicts that dissipative disorder gives rise to an unexpected super-diffusive spatial transport, which is characterized by ultra-far spatial jumps between localized states (Fig. 1c). We implement our ideas in photonic lattices consisting of coupled fibre loops and experimentally demonstrate the existence of far jumps between the localized eigenstates as well as the super-diffusive transport in systems with stochastic dissipation.

**Analytical model**

We start by developing an analytic model for wave transport in a system with stochastic dissipation. Our approach considers a general lattice with $\psi_n$ as the field amplitude on site $n$. In his model, Anderson considered real lattice site potentials, i.e. the Hermitian case. He found that for a random potential all eigenstates can become exponentially localized and consequently, spatial transport comes to complete halt. In contrast, we consider complex potentials and random fluctuations in the imaginary part of the potential, that is, stochastic dissipation in the lattice. As it was shown before, also in this case all eigenstates are exponentially localized[13].

However, in stark contrast to the Hermitian case, the dissipative character of the lattice causes the population of the eigenstates to change during evolution. Consequently, the mere fact that all eigenstates are localized does not necessarily cause the suppression of transport. We show this by considering the eigenvalue spectrum of our system. After exciting the lattice with a single-site excitation, the population of the modes is determined by weight factors that are governed by the overlap between the initial single site excitation and the exponentially localized states. Hence, a weight factor is proportional to $\exp(-d/l_c)$, where $d$ is the distance between the excited site and the centre of the state, and $l_c$ the localization length of the exponentially localized state (Fig. 2a). Moreover, due to the random character of the lattice the weight factors of each eigenstate $j$ experience a phase oscillation of the form $\exp(i\text{Re}(\lambda_j)t)$, where $\lambda_j$ is the eigenvalue of the corresponding state and $t$ the time. As all eigenvalues of the eigenstates possess different imaginary parts, the weight factor of each mode is also modified by a term $\exp(-\text{Im}(\lambda_j)t)$, such that its population, relative to other eigenstate, changes over time. Let us assume that at time $t = 0$ primarily the eigenstate with eigenvalue $\lambda_0 \in C$ is excited, whereas the weight factor of a another state, with eigenvalue $\lambda_d$ and localization centre at position $d$, is smaller by the factor $\exp(-d/l_c)$. If $\text{Im}(\lambda_d) > \text{Im}(\lambda_0)$, then the state with eigenvalue $\lambda_d$ will become dominant at a time $t_{\text{jump}}$, which is given by

$$\exp(-d/l_c)\exp((\text{Im}(\lambda_d) - \text{Im}(\lambda_0))t_{\text{jump}}) \approx 1. \qquad (1)$$

At this point, for $t > t_{\text{jump}}$ the weight factor of the state with eigenvalue $\lambda_d$ exceeds the one of the initially excited state with eigenvalue $\lambda_0$, such that the population jumps from the excited state to the second one in spite of the localization. This is sketched in Fig. 2a; these jumps form the basis for the type of spatial "Anderson transport" we are discussing here.

For strong disorder, our model allows to analytically quantify this disorder-induced transport. To this end, we compute the probability $P_n(t)$ that the population has jumped from an initial lattice site $n = 0$ to a site $n$ at time $t$. Due to the stochastic dissipation, the imaginary parts of the eigenvalues are randomly distributed and follow the distribution function $f(\rho)$. The probability for a jump to position $n$, where the state has an imaginary part of the eigenvalue in the range $(\rho, \rho + d\rho)$, is then determined by the joint probabilities that all other states $j \neq n$ are not dominant yet, hence $\text{Im}(\lambda_j)t - |j|/l_c < \rho t - |n|/l_c$. Here, we assume that the imaginary parts of the eigenvalues of different modes are independent stochastic variables with the same distribution $f(\rho)$. This assumption is particularly valid as long as the eigenstates extent only over a few sites, as it is the case for strong disorder. The jump probability can then be calculated via

$$P_n(t) = \int_{\frac{|n|}{tl_c}}^{\infty} d\rho\, f(\rho) \prod_{j=0,\pm 1,\pm 2,\pm 3\ldots \neq n} \int_{-\infty}^{\rho + \frac{|j|-|n|}{tl_c}} d\xi\, f(\xi) \qquad (2)$$

The product of the inner integrals represents the probability that all other modes $j \neq n$ have a decay rate smaller than $\rho$. The outer integral takes into account the probability that the considered mode at distance $n$ and time $t$ has an eigenvalue $\rho$ that is large enough to observe the jump. Note that $P_n(t)$ and, hence, the transport is mainly determined by the distribution function $f(\rho)$. Importantly, we did not use a specific model for the derivation of Eq. (2); the only condition is that the system under consideration has to exhibit localized eigenstates.

As a common quantity to characterize transport, we consider the time evolution of the second moment $M_2(t) = \sum_n n^2 |\psi_n(t)|^2$ for a wave packet initially localized at $n = 0$, where $|\psi_n(t)|^2$ is the normalized population at site $n$. The time evolution of the average second moment can then be calculated via $\langle M_2(t) \rangle = \sum_n n^2 P_n$, with $\langle \cdot \rangle$ denoting an averaging over different random disorder realizations. Its slope $s = \frac{d \log \langle M_2(t) \rangle}{d \log t}$ is a general measure of the transport speed. For example, ballistic transport is characterized by $s = 2$, whereas for diffusion one finds $s = 1$. For sufficiently strong disorder, such that our assumptions hold, our analytical model yields a slope of $s \to 4/3$, which is super-diffusive (Fig. 2b). The exact value of s is determined by the distribution function $f(\rho)$, which, in turn, depends on the exact experimental implementation (see Supplementary Information), such that for strong disorder one can obtain a triangular distribution for which we find $s_\triangle \to 4/3$ but also a rectangular distribution for which we get $s_\square \to 1$. The analytical results are in excellent agreement with the numerical evaluation of Eqs. (3), which describe our experimental platform and will be introduced below. This agreement is remarkable, especially because the analytic model and the numerical propagation are two completely independent ways to obtain *s*. Moreover, for weak disorder the numerical data suggest that the transport speed *s* increases and reaches the ballistic regime when dissipation ceases. Hence, Anderson transport is super-diffusive, independently of the disorder strength.

In the Supplementary Information, we show that for strong disorder our analytical model can not only predict the growth *s*, but - for suitable localization length $l_c$ - also the absolute values of $\langle M_2(t) \rangle$. Additionally, in order to extend our theoretical basis, we provide a rigorous analytical derivation of the exponential localization of eigenstates and spreading dynamics in media with stochastic dissipation for the special case of the Hatano-Nelson model[14], which displays the same kind of non-Hermitian transport in the localized phase.

**Experimental implementation**
Now we turn to the experimental demonstration of our theoretical findings. For our studies, we employ classical light propagation in coupled optical fibre loops[30,31]. They form a one-dimensional photonic lattice with precisely tuneable parameters as a model for the evolution of single quantum particles and wave transport in a variety of lattice systems[32–35]. The ability to adjust the strength of dissipation at will enables the implementation of nearly arbitrary complex potentials, especially random dissipation. To study localization and transport, we evaluate the light propagation that arises from single-site excitations of the lattice, which corresponds to the evolution of the probability density of an electron that initially resides at a specific atom.

The working principle of our experimental platform is to let optical pulses propagate in a pair of unequally long fibre loops, which are connected by a beam splitter (Fig. 3a). A detailed discussion of the full setting is presented in the Methods section. The pulse dynamics in the loops can be mapped onto the light evolution in a 1+1-dimensional double-discrete lattice, as shown in Fig. 3b and discussed in the Supplementary Information. The light evolution is governed by a set of coupled equations[31]

$$u_n^{m+1} = \frac{G_u}{\sqrt{2}}(u_{n+1}^m + iv_{n+1}^m)e^{i\varphi_u}$$

$$v_n^{m+1} = \frac{G_v}{\sqrt{2}}(iu_{n-1}^m + v_{n-1}^m)$$

(3)

where $u_n^m$ denotes the amplitude at lattice position $n$ and time step $m$, on left moving paths, and $v_n^m$ the corresponding amplitude on right moving paths. The quantities $\varphi_u$ and $G_{u,v} = G_{u,v}(n,m)$ are degrees of freedom to control the imaginary and the real part of the lattice potential, respectively. As such, our platform can model Hermitian and non-Hermitian disorder by choosing either $\varphi_u$ or $G_{u,v}$ as a random variable.

To study localization and transport, we evaluate the propagation that arises from a single-site excitation, which can be either $v_n^0 = \delta_{0n}$ or $u_n^0 = \delta_{0n}$. The squared modulus of the wave function $|\psi_n(t)|^2$ is represented by the normalized light intensity distribution $|u_n(t)|^2 + |v_n(t)|^2$ within the lattice. The time $t$ corresponds to the time step $m$, i.e. $u_n(t) = u_n^m$. For the characterization of the localization and transport features, it is sufficient to use the light intensity distribution of one fibre loop, as we confirmed by numerical analysis, since the differences between the loops are only on a local scale and these vanish upon averaging. For this reason all experimental data and the corresponding numerical data are based on the intensity distribution in the $u$-loop.

**Experimental results**

In the homogeneous, disorder-free lattice $G_u = G_v = 1$, $\varphi_u = 0$, a single-site excitation yields the well-known ballistic spreading of the wave function[23] (Fig. 3c left). As a result, the initially localized wave packet quickly becomes delocalized and acquires a high probability to be found far away from its initial position. Now we consider the two cases with disorder. In the conventional Hermitian case, disorder is commonly realized by a time-independent but spatially random real part of the potential[36]. We realize the random changes in the real part of the potential by drawing $\varphi_u \in [-W, W]$ for each lattice site $n$ from a uniform probability distribution with disorder strength $W$. In accordance to previous studies[37,38], a single-site excitation in such disordered lattice undergoes repeated scattering at the potential fluctuations, leading to a superposition of destructively interfering waves in such a way that previously extended states localize at the initial position (Fig. 3c centre). This process is precisely the Anderson localization.

Within the non-Hermitian Anderson model, disorder comes into play by a stochastic energy exchange with the environment, i.e. random changes in the imaginary part of the potential. The imaginary potential fluctuations are given by

$$G_{u,v} = e^{i\gamma}, \ \gamma \in [-iW, iW] \qquad (4)$$

and are also drawn from a uniform probability distribution for each lattice site. While this distribution also includes gain, the results do not depend on whether a purely passive system or a system with gain and loss is chosen, which has been confirmed in earlier theoretical works[13]. We find, numerically and in the experiments, that a single-site excitation localizes due to the non-Hermitian disorder (Fig. 3c right), which is in agreement to the predicted localization of eigenstates[13]. Yet, we observe the spatial transport via transverse jumps, in agreement with the prediction of our analytical model. Remarkably, the distances spanned by these jumps can significantly exceed the localization length of the eigenstates. This demonstrates that in non-Hermitian systems even extremely small overlaps between the localized eigenstates can suffice to substantially influence the temporal light evolution. Moreover, not only the average, but also the individual experimental light evolutions agree very well with the numerical propagation of Eqs. (3) (see Supplementary Information).

To further quantify the observed transport, we evaluate the time evolution of the second moment $M_2(t)$ for an excitation initially localized at $n = 0$. Here, $M_2(t)$ is a random variable that changes with every disorder realization. Consequently, the data which are extracted from the experiments, contain the second moments of different disorder realizations, from which we then derive the mean $\langle M_2(t) \rangle$ and the standard deviation $\sigma(M_2(t))$. These quantities allow to draw statistically meaningful conclusions for the transport. In the Supplementary Information, we also evaluate the statistical movement of the wave function's centre of mass, in order to further characterize the mode jump transport.

Our experimental results (Fig. 4) clearly demonstrate that the transport behaviour of the non-Hermitian disorder fundamentally differs from the Hermitian case, even though in both cases all eigenstates are exponentially localized. Furthermore, the experimental data successfully match the numerical results obtained from Eqs. (3) within the range of one standard deviation of expected statistical fluctuations. Our main findings can be directly concluded from the experimental data: First, the dissipative disorder enables the wave function to be possibly found everywhere in the lattice after a finite period of time, which is in strong contrast to the Hermitian case, where the wave packet is predetermined to remain arrested forever at its initial position. This conclusion follows from the monotonous increase of $\langle M_2(t) \rangle$ in time. In contrast, in the Hermitian case, the mean of the second moment quickly saturates. The second main finding concerns the transport speed, as it demonstrates that random dissipation indeed induces a super-diffusive transport.

These results pose the question of what happens when both, the real and the imaginary part of the potential are subject to random changes. Disorder in the real part of the potential leads to Anderson localization and completely suppresses transport, whereas disorder in the imaginary part of the potential induces super-diffusive transport. Our experimental data unequivocally show that Anderson localization cannot suppress the non-Hermitian mode jumping transport as the stochastic energy exchange with the environment still facilitates a transverse energy flow. These results are summarized in the corresponding section of the Supplementary Material.

**Conclusion**

We have observed the disorder-induced coexistence of a fully localized eigenstate spectrum and a super-diffusive wave transport. This non-Hermitian "Anderson transport" is driven by a stochastic energy exchange with the environment. All of our predictions could be successfully demonstrated in our experiments. Since our analytic model solely relies on an eigenstate analysis, our results can be applied in principle to numerous platforms beyond photonics, such as matter waves, acoustic waves, and electrons. As every real-world system is subject to dissipation and disorder, we anticipate our findings to inspire an extended understanding and controlling of localization and transport properties in a plethora of classical and quantum physical systems.

**Figures**

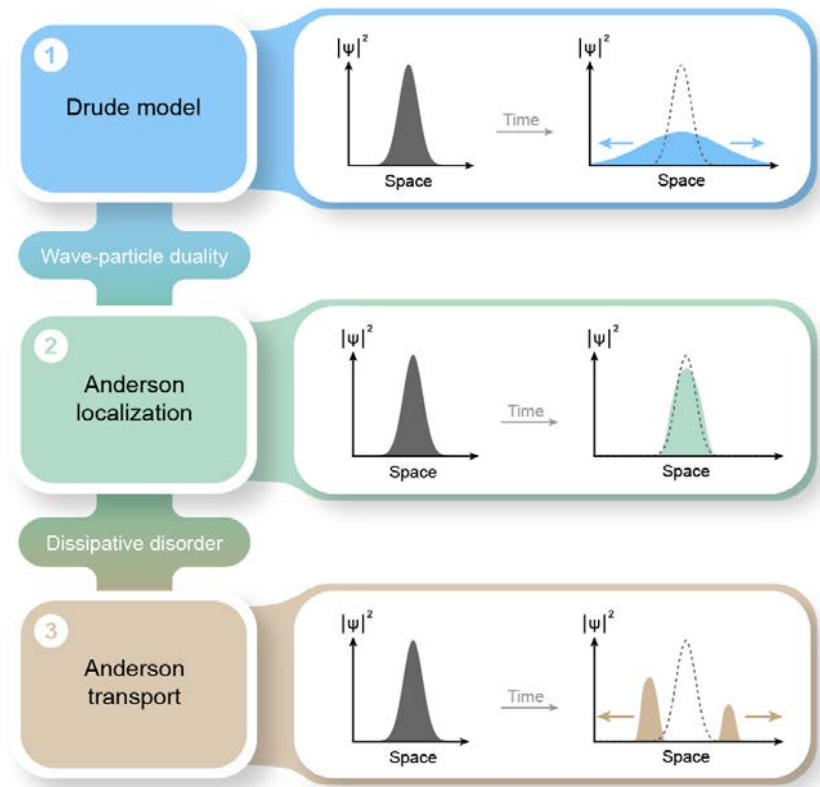

**Figure 1 | Electronic transport in disordered lattices**. The road from the classical Drude model to the quantum mechanical Anderson model and to Anderson transport. Within the Drude model (top) electrons are considered as particles and move in a disordered lattice via random collisions, which statistically yields a diffusive spreading of the probability density $|\psi|^2$ in space and time, according to the Born rule. When taking the wave nature of electrons into account, in disordered lattices the electron movement will be fully suppressed due to Anderson localization (centre). The generalization to dissipative (i.e. non-Hermitian) disorder will suddenly facilitate "Anderson transport", which is caused by spatial jumps between localized states (bottom).

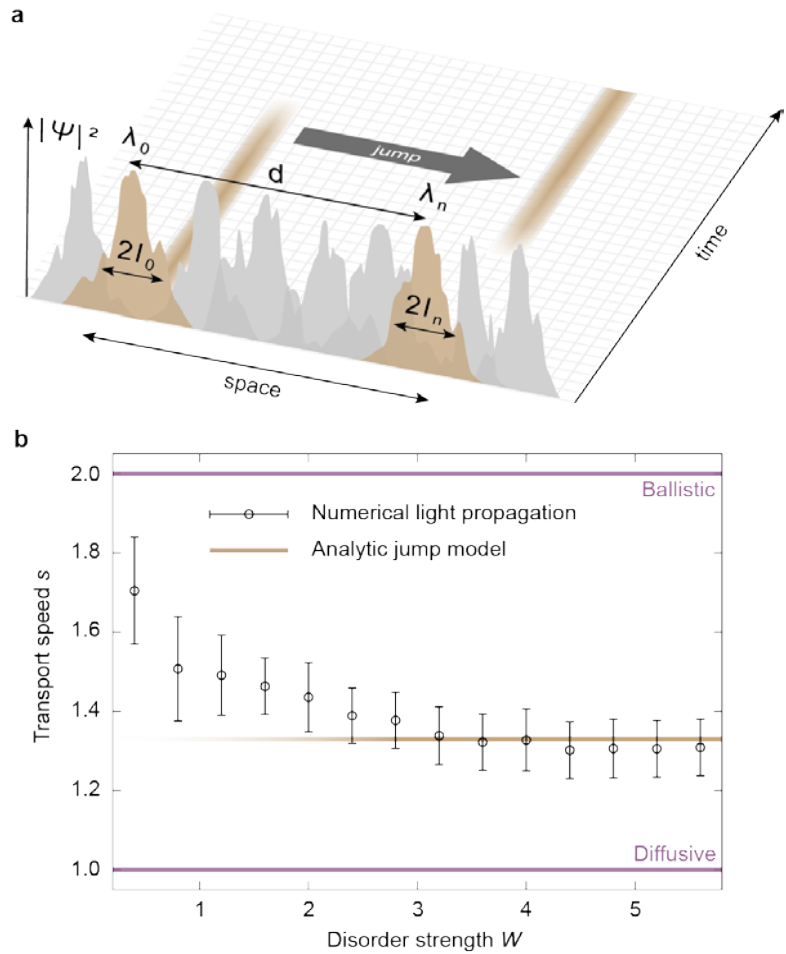

**Figure 2 | Theory of Anderson transport. a**, Eigenstate jump scheme of the analytic model. The exponentially localized eigenstates of a lattice with dissipative disorder exhibit eigenvalues $\lambda$ with different imaginary parts. Initially, the left state is dominant. If $\text{Im}(\lambda_n) > \text{Im}(\lambda_0)$ the state at site $n$ will become dominant at some time $t_{\text{jump}}$, which depends on the localization lengths $l_0, l_n$ and the distance $d$ of the two states. Around $t_{\text{jump}}$, there is a rapid spatial transition of the population from the left to the right state, appearing as a "jump". **b**, Analytically, Eq.(2), and numerically, Eqs. (3-4), extracted transport speed $s(W)$ as a function of the disorder strength $W$. Our analytical predictions agree very well with the numerical data for strong disorder. The error bars (see Supplementary Information) capture one standard deviation of $s(W)$. For weaker disorder, our numerical data suggest that the transport disorder speed increases as the eigenstates become less localized.

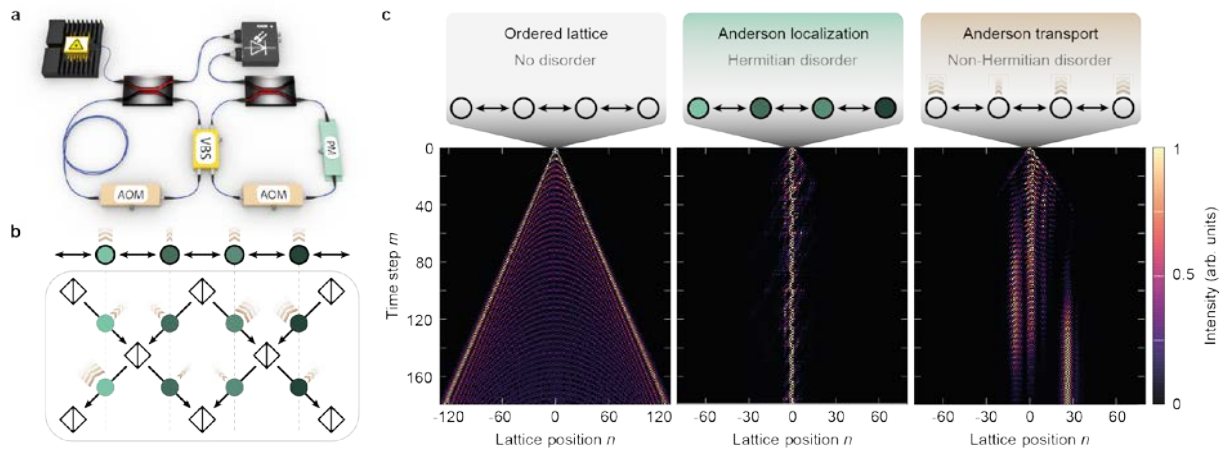

**Figure 3 | Experimental implementation of Anderson transport**. **a**, Simplified experimental arrangement for temporally encoded lattices. Two unequally long optical fibre loops are connected by a variable beam splitter (VBS), which is set to 50:50 splitting. Acousto-optical modulators (AOM) control the dissipation. A phase modulator (PM) controls the on-site energies. A photo detector measures the temporal light evolution. **b**, A linear chain of coupled sites (top) is realized with a 1D quantum walk (bottom). Different arrow widths correspond to different dissipation strengths. Different shades of green correspond to different on-site energies. **c**, The experimental propagation through our lattice shows ballistic spreading for homogeneous lattices (left) compared to Anderson localization in Hermitian disordered lattice (centre) and the jumping between localized states in the case of Anderson transport in disordered dissipative lattices (right).

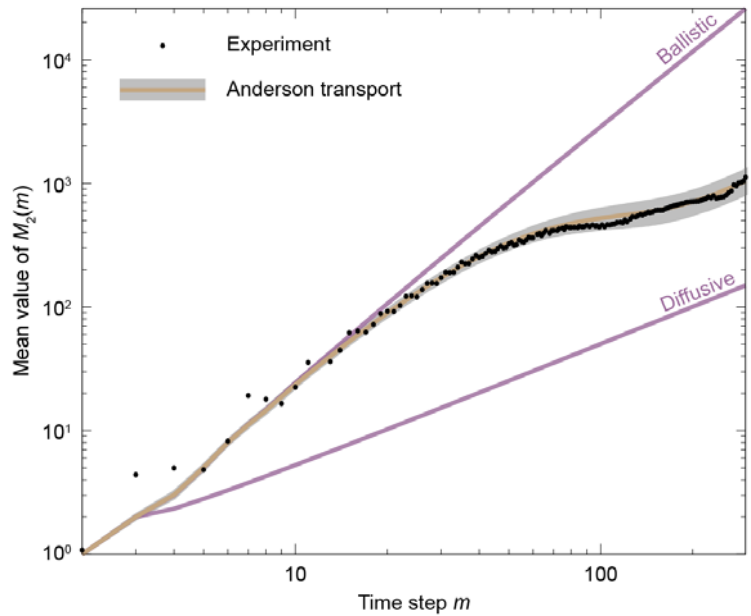

**Figure 4 | Transport by dissipative disorder.** Time evolution of the mean value of the second moment $M_2(t)$. The experimental data are based on 40 random disorder realizations of dissipation. For a large number of disorder realizations, the numerical data converge to the shown line for Anderson transport. The grey area shows plus-minus one numerical standard deviation of expected statistical fluctuations, which arise from using only 40 disorder realizations instead of a very large number. For comparison, ballistic and diffusive transport are shown.

**Methods**

**Experimental setup**

A single 50ns pulse is launched into a fibre loop, which is coupled to another fibre loop by a 50:50 beam splitter (Fig. 3a). While the pulse circulates in the fibre loop arrangement it repeatedly splits up at the beam splitter and multipath interference between the emerging sub-pulses takes place. Each loop contains an acousto-optical modulator (Brimrose Corp.) to manipulate the amplitude of propagating pulses, which allows to realize dissipative lattices, by changing the transmission of the modulator in time. To avoid shifts in the frequency, the $0^{th}$ diffraction order is aligned to the output. Fused fibre coupler (AC Photonics) are used to inject light into the loops and to couple parts of the light out to photodetectors (Thorlabs). Optical isolators (AC Photonics) suppress back reflections and ensure unidirectional propagation. Spools of single mode fibre (Corning® Vascade® LEAF® EP) extend the propagation time for each loop to approximately 27µs. A fibre optic patch cord is added in one loop and induces a 100ns difference in propagation time, such that fibre arrangement features a long and short loop. The shorter loop contains a phase modulator (iXblue Photonics) which allows manipulating the phase of propagating pulses. Each loop contains an erbium-doped fibre amplifier (Thorlabs) to compensate for propagation losses that have no physical meaning for the propagation in the synthetic lattice, e.g. insertion losses and detection losses. The amplification-scheme is balanced between both loops and does not alter the multipath interference. The amplification allows to maintain a high signal-to-noise ratio over many loop round trips. The erbium-doped fibre amplifiers are optically gain clamped with a distributed feedback laser (JDS Uniphase, 1538nm), which is coupled to the amplifiers by a wavelength division multiplexing coupler (AC Photonics). Optical band-pass filter (WL Photonics) remove the gain clamping light and suppress amplified spontaneous emission from the amplification process. All fibre components are designed for operation at 1550nm wavelength and employ standard single mode fibre, e.g. SMF28 or comparable. Mechanical polarization controller (Thorlabs) are used to align the polarization state in front of polarization-sensitive components. The input signal is a 50ns rectangular shaped pulse, which is cut out of a continuous wave signal of a DFB laser diode (JDS Uniphase, 1550nm) with a Mach-Zehnder intensity modulator (SDL Integrated Optics Limited). To obtain a higher on-off ratio, another acousto-optical modulator is used to only transmit the seed pulse and further decrease the off-level. Arbitrary waveform generators (Keysight Technologies, 33622A) perform the signal generation for all electro-optical devices.

**Data acquisition** To measure the statistical behaviour of Anderson transport, 40 random realizations of dissipative disorder were generated in MATLAB and then converted into a waveform signal, which is applied to the amplitude modulators by the arbitrary waveform generators. To obtain the squared modulus of the lattice site amplitudes in the presented data sets, we performed a time-resolved measurement of pulse intensities with a photodiode (Thorlabs) in the shorter fibre loop. The output voltages of the photodiode are amplified with a logarithmic amplifier (FEMTO HLVA-100) and afterwards sampled with an oscilloscope (R&S RTO1104). With the time scales $\Delta t$ and $T$ of the fibre loop arrangement, one can map the acquired voltage signal onto the discrete 1+1D grid $(m, n)$ in which the measured pulse intensities represent the squared modulus of the lattice site amplitude for the respective propagation step $m$ and lattice position $n$. Each measurement features an additional noise measurement in which the lattice excitation is turned off. In the post-processing, this noise data are subtracted from the original data in which the lattice was excited. Based on the post-processed data, the temporal evolution of the first and second moment is evaluated.

**Acknowledgements**

The authors thank the Deutsche Forschungsgemeinschaft for funding their research (grant SZ 276/20-1) and the Krupp von-Bohlen-and-Halbach foundation.

**Author Contributions**

SW and MK designed the experimental implementation. SW performed the experiments. SL developed the analytical model. AS supervised the project. All authors discussed the results and co-wrote the manuscript.

**Competing Interest Declaration**

The authors declare no competing interests.

**Additional Information**

Supplementary Information is available for this paper. The Supplementary Information contains details on the temporal encoding of the photonic lattices. Further experimental results are shown. Aspects related to the numerical transport evaluation are detailed. A rigorous analytical derivation of the localization of eigenmodes, their localizations lengths and a solution for Anderson transport in the unidirectional Hatano-Nelson model is provided.

Correspondence and requests for materials should be addressed to alexander.szameit@uni-rostock.de.